# Design, Modelling, and Simulation analysis of a Single Axis MEMS-based Capacitive Accelerometer

Veena. S[1], Newton Rai [2], H.L. Suresh[3], Veda Sandeep Nagaraja[4]

[1]*Assistant Professor, Department of Electrical and Electronics Engineering, Nitte Meenakshi Institute of Technology, Bangalore, Karnataka, India;veena.s@nmit.ac.in*
[2]*Department of Electrical and Electronics Engineering, Nitte Meenakshi Institute of Technology, Bangalore*
[3]*Professor and Head, Department of Electrical and Electronics Engineering, Sir. M Visvesvaraya Institute of Technology, Bangalore*
[4]*Senior Scientist, Tyndall National Institute, University College Cork, Ireland*

*Abstract* - This paper presents the design, simulation, and analytical modeling of the single proposed axis MEMS-based capacitive accelerometer. Analytical modeling has been done for frequency and displacement sensitivity. The performance of the accelerometer was tested for both static and dynamic conditions, and the corresponding static capacitance value was calculated and was found to be $C_0=0.730455pF$, a response time of 95.17μs, and settling time of 7.261ms and the displacement sensitivity $S_d = 3.5362*10^{-8}$ m/g. It was observed that the sensitivity of the accelerometer depends on its design parameters like beam length, overlap area of comb, sensing mass, and the number of interdigital fingers. A novel capacitive accelerometer has been designed for an operating frequency of 2.1kHz

The accelerometer was designed using COMSOL Multiphysics and analyzed using the MATLAB simulator tool. The single proposed axis MEMS-based capacitive accelerometer is suitable for automobile applications such as airbag deployment and navigation.

## I. INTRODUCTION

The accelerometer is an electromechanical device that is used for physical measurements like acceleration, force, the vibration of a moving solid. Micro machined accelerometers are one of the important classes of MEMS devices. There is a wide scope of utilization that requires acceleration measurement such as automotive industry, biomedical applications, oil and gas exploration, vibration analysis, navigation system, robotics, mobile, and computer accessories.

Most accelerometers are based on the principle of mechanical vibration. The fundamental structure of the MEMS accelerometer contains the seismic mass supported by beams. The mass is frequently appended to a dashpot that gives the essential damping impacts [1,2]. The spring and the dashpot are in turn connected to the frame, as shown in figure 1. An accelerometer that is kept at rest will measure acceleration due to gravity of g ≈ 9.81 m/s², and in contrast, accelerometers that are in a free-fall state will measure as zero [3].

Accelerometers are classified based on their principle of operation. They are piezoelectric, piezoresistive, capacitive, heat transfer, optical, hall effect, thermal, interferometric, etc. [4], but commercially, piezoelectric, piezoresistive, capacitive were widely used.

**Piezoelectric:** utilizes piezo ceramics like lead zirconate titanate, and they have a very high-frequency range, large measuring range up to 6000g, self-powered device [5].

**Piezoresistive:** employ beam-like structure whose resistance varies with acceleration. They are cheap due to their simple construction design, low hysteresis, simple readout circuits and have the ability to operate at higher temperatures [6].

**Capacitive:** the capacitive-based MEMS accelerometers measure capacitance change between a fixed and movable electrode isolated by a littlegap[7].

Although commercially different accelerometers are available, the main aim for selecting capacitive based MEMS accelerometers are their high Sensitivity and linearity, good repeatability, low noise performance, flexible structure to design, uses low power to operate, and high durability

This paper presents a novel single folded beam-type capacitive MEMS Accelerometer. A literature survey is done and is presented in section II. Principle of operation and the design of the proposed accelerometer is discussed in section III and IV respectively. Mathematical modeling of the proposed accelerometer is presented in section VI and followed by the simulation results. An open-loop model of the proposed accelerometer and the results are discussed in consecutive sections.

## II. LITERATURE REVIEW

Lots of work has been done in the field of MEMS comb-type accelerometers, and numbers of publications are available. Many authors have proposed their ideas on design, working, mathematical analysis of MEMS accelerometer and still working on improving the performance of the same.

Babak VakiliAmini et al. [8] have discussed details on air squeeze film damping and analytical formula for calculation of damping coefficient and discussed the mathematical model for an open and closed-loop model of capacitive accelerometers. This paper suggests an idea about the calculation of damping coefficient efficiently and an idea of the second-order open and closed-loop operation of MEMS accelerometers.

W. WAI-CHI et al. [9] discussed various methods for the calculation of stiffness constant for a folded beam structure. The stiffness value depends on the structural design, like beam length, width, and thickness. The stiffness plays a vital role in the accelerometer specification. The author suggested the formulation of the stiffness constant from theoretical analysis.

Tulika et al. [10] have discussed the mathematical model of the comb-type capacitive accelerometer, as well as sensitivity analysis of interdigital comb-type structure. The author suggested the interdigital comb, improvement of sensitivity by optimizing design parameters of MEMS accelerometer.

From the above references, a novel single axis, single folded beam-type capacitive MEMS Accelerometer has been designed.

### III. PRINCIPLE OF OPERATION

The fundamental principle of working a MEMS accelerometer is the displacement of a proof mass suspended by springs. The accelerometer structure consists of a proof mass attached to a fixed frame through springs having a spring constant (k) and often damper having a damping coefficient (b) as shown in figure 1, where the damper provides the necessary damping effects [12-13].

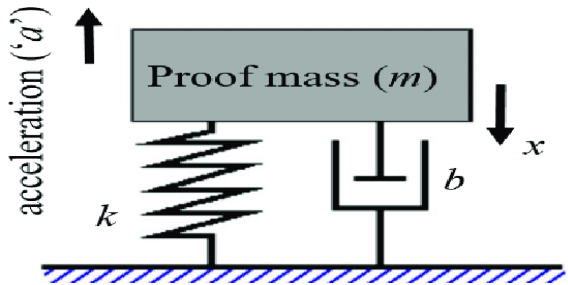

**Fig 1: Typical spring mass damper model of accelerometer**

When an external inertial force is applied, the proof mass moves with respect to the reference frame, which results in stress in the suspension spring. The relative displacement of proof mass with respect to frame and stress caused in spring gives the measurement of the external acceleration [14].

The movement of proof mass results in a change in displacement between interdigital fingers, so this displacement causes a change in capacitance between interdigital comb fingers. The capacitance change can be sensed by interfacing with a capacitive readout circuit in order to measure the acceleration. The accelerometer can be modeled as a second-order spring mass damper system [15].

Let 'x' be the displacement of proof mass 'M' relative to the frame and 'a' be the external acceleration. Then, the Force is expressed as

$$F_{total} = F_{inertial} + F_{damping} + F_{spring} \text{-(1)}$$

Where, $F_{inertial} = -Ma$, $F_{damping} = bx$, $F_{spring} = Kx$

The dynamic behavior of this system is governed by Newton's second law of motion [16,17].

$$\frac{m dx^2}{dt^2} + \frac{b dx}{dt^2} + kx = F = ma \text{-(2)}$$

### III. ACCELEROMETER DESIGN

The structure design of the proposed comb-type MEMS accelerometer is shown in figure 2. The proposed design of the MEMS accelerometer has a unique structure where two proof masses were suspended by a single folded beam. The structure of the proposed design has symmetry in shape. It consists of movable and fixed parts. Movable parts comprise two proof mass, single folded beam, movable fingers, and fixed parts comprise anchors, pad metal, fixed fingers attached to pad metal [18]. The two proof masses are symmetrically suspended to the central anchor through a single folded beam. The fixed fingers are connected to the left and right anchors.

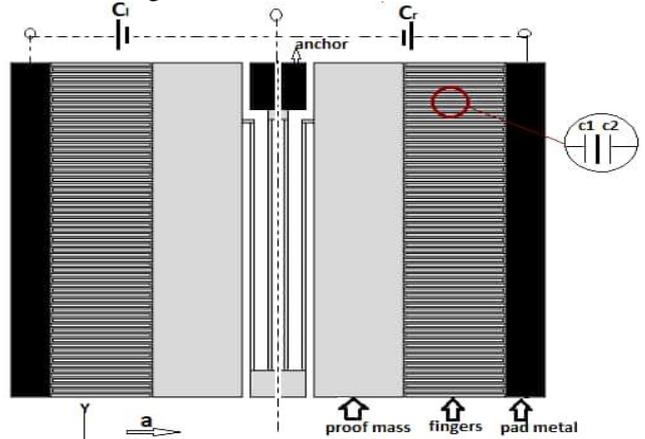

**Fig 2: Structure design of proposed MEMS comb accelerometer**

The two proof masses are responsible for vibrating in response to the incident inertial force. The movable fingers are attached on one side of each proof mass, whereas the other sides of the proof mass are attached to the beam. The movable fingers are surrounded by fixed fingers which constitute the differential capacitance pair $C_1$ and $C_2$ with fixed fingers on either side [19]. The structure consists of two positive electrodes and one ground electrode [20]. The number of movable fingers ($N_f$) and fixed fingers are 66 and 68, respectively. Table I gives the dimensions of the proposed accelerometer.

**TABLE I: DIMENSIONS OF MEMS ACCELEROMETER**

| Accelerometer | Dimensions (µm) |
|---|---|
| Right proof mass | 225X1000X25 |
| Left proof mass | 225X1000X25 |
| The gap between two fingers ($d_0$) | 5 |

| Pad metal size | 100X1000X25 |
|---|---|
| Thickness of device (t) | 25 |
| Beam length (L) | 250 |
| Beam width (W) | 10 |
| Length of the finger ($l_f$) | 245 |
| The breadth of the finger ($b_f$) | 10 |

## IV. WORKING PHENOMENON

The working phenomenon of Single-axis MEMS comb finger type accelerometer is as shown in the figure 3 and 4

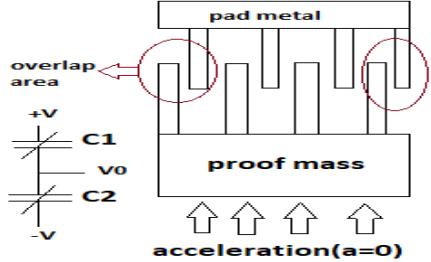

Fig 3: Schematic of fingers displacement when a=0

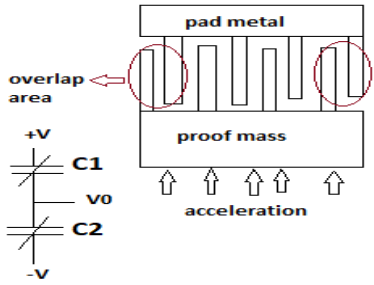

Fig 4: Schematic of fingers displacement when a ≠ 0

At steady state (when acceleration=0), the proof mass doesn't experience any inertial force, so the movable fingers are in rest. As there is no change in area between fixed and movable fingers so the leftcapacitance $C_1$ is equal to right capacitance $C_2$.
Under steady state an external driving voltage of 5V dc is applied through positive and negative electrodes, which develops an electrostatic force, and this force will try to displace the sensing electrodes and the corresponding capacitance under this condition is called static capacitance. The static capacitance is expressed as

$$C_0 = \frac{EN_f l_f t}{d_o} \text{-(3)}$$

Where, $C_0$ is the static capacitance.
When an external acceleration is applied in the direction along horizontal x-axis parallel to the reference frame,themovable mass (proof mass+movable fingers) experiences an inertial force and is displaced in opposite direction to the applied inertial force. Since, the sensing fingers moves with a certain displacement say'x' which results in change in overlap area between fixed and movable fingers. Hence the differential capacitances $C_1$ is no longer equal to $C_2$ ($C_1 \neq C_2$) and will have some specific value. The parallel plate capacitor is formed between moving and fixed fingers and air acts as dielectric medium. The value and direction of force applied can be known by measuring the change in differential capacitance on both left and right parts of accelerometer. The value of capacitance $C_1$ and $C_2$ is given by the equation [21].

$$C_1 = \frac{\varepsilon N_f(x_1+x)t}{d_o} \text{-(4)}$$

$$C_2 = \frac{\varepsilon N_f(x_1-x)t}{d_o} \text{-(5)}$$

Where, $x_1$ is the initial overlap length between fixed and moving fingers, ε is permittivity of air, C1 and C2 are change in capacitance.

## V. MATHEMATICAL MODELLING OF ACCELEROMETER

As inertial force is dominant over all other forces acting on the proof mass, equation 2 can be written as

$$\frac{m dx^2}{dt^2} + \frac{b dx}{dt^2} + kx = F = ma \quad \text{-(6)}$$

Hence on applying Laplace transform to equation 6,
We get, $\frac{X(s)}{a(s)} = \frac{1}{s^2 + \frac{b}{m}s + \frac{k}{m}}$ -(7)

The above equation (7) can be rewritten in a standard form with ζ and $w_n$ as

$$\frac{X(s)}{a(s)} = \frac{1}{s^2 + 2\zeta s w_n + w_n^2} \text{- (8)}$$

Where $W_n$ is the natural frequency in rad/sec and ζ is the damping ratio.

### A. To Find the Natural Frequency:
The natural frequency of a spring mass damper system is

$$w_n = \sqrt{\frac{K}{m}} \quad \text{or} \quad f_n = \frac{1}{2\pi}\sqrt{\frac{k}{m}} \text{-(9)}$$

Where, $W_n$ is the natural frequency in rad/sec and $f_n$ is the natural frequency in Hertz, m is the total mass and k is the stiffness of the spring.
Here the total sensing mass of accelerometer [22] is given by the sum of proof mass and the moving finger mass

### B. To Find the Mechanical Stiffness (K):
The folded beam suspension is way better than any other beam in terms of sensitivity. The folded beam has lower value and is more stiffer than any other beams. In this design there are totally four beams, stiffness of spring in parallel add up whereas stiffness of spring in series add up reciprocally.
The spring constant for the folded beam is given by [9]

$K = \frac{EtW^3}{4L^3}$ - (10) Where L is the length of beam, t is the height of beam, W is the width of beam, E is the Youngs modulus of silicon=$170*10^9 N/m^2$

### C. Displacement Analysis:
Referring to equation 6, it is understood that under steady state the derivative of x becomes zero. Hence the equation reduces to ma=Kx -(11)
The displacement of proof mass (x) is now given by

$$x = \frac{m}{K} * a \quad \text{-(12)}$$

On rearranging the above equation, we get the displacement sensitivity of proof mass as

$$S_d = \frac{X}{a} = \frac{m}{K} \quad \text{-(13)}$$

### D. To Find the Damping Ratio ($\zeta$):

The nature of the system response directly depends upon the value of damping ratio ($\zeta$). The dynamic analysis of the system is done to find the damping ratio and hence the behaviour of the system is analysed.

- *Squeeze Film Damping:*

The pressure is developed due to the compression of air between moving fingers and fixed fingers and this effect is called damping effect. It acts as a dragging force and opposes the movement of sensing fingers [23]. Equation 14 gives the analytical formula for the air squeeze film damping[8].

$$b = N_f n_{eff} l_f \left(\frac{t}{d_o}\right)^3 \text{-(14)}$$

Where, $n_{eff}$ is the effective viscosity of air which is given by $18.5*10^{-6}$ Ns/m$^2$

- *At Dynamic State:*

The dynamic behaviour of spring mass damper system is defined by parameters like angular frequency ($w_n$) and damping ratio($\zeta$). The dynamic analysis of this system is done in Laplace domain of equation (2) as shown below [16].

when $x \to F$

$$\frac{X(s)}{F(s)} = \frac{1}{ms^2+bs+k} \text{-(15)}$$

when $x \to a$

$$\frac{X(s)}{a(s)} = \frac{1}{s^2+\frac{b}{m}s+\frac{k}{m}} \text{-(16)}$$

On comparing equation 17 with equation 8, we get the damping ratio as

$$\text{Damping ratio}(\zeta) = \frac{b}{2mw_n} \text{-(17)}$$

The actual theoretical calculations for the parameters such as natural frequency, damping coefficient, displacement of proof mass etc of the proposed accelerometer are obtained from the equations 9 to 17 and are tabulated in Table II.

**TABLE II: ANALYTICAL VALUES**

| Parameters | Calculated Values |
|---|---|
| Static capacitance ($C_0$) | 0.730455pF |
| Mass (m) | $3.53625*10^{-8}$Kg |
| Spring constant (k) | 10 N/m |
| Natural frequency ($f_n$) | 2.6 kHz |
| Displacement of proof mass(x) | 0.0353625μm |
| Displacement sensitivity of proof mass ($S_d$) | $3.53625*10^{-9}$ m/g |
| Damping coefficient(b) | $3.815625*10^{-5}$Ns/m |
| Damping ratio($\zeta$) | 0.03208 |

From Table II it is observed that the damping ratio for this system is less than one ($\zeta < 1$), hence the proposed accelerometer is an underdamped system.
For the underdamped system,

$$\text{Rise time }(t_r) = \frac{\pi - \tan^{-1}\frac{\sqrt{1-\zeta^2}}{\zeta}}{w_n\sqrt{1-\zeta^2}} \text{- (18)}$$

$$\text{Settling time }(t_s) = \frac{4}{\zeta w_n} \text{- (19)}$$

The values obtained for $t_r$ and $t_s$ are
$t_r$= 95.36 μs  and $t_s$=7.583 ms

## VI. SIMULATION RESULTS

COMSOL Multiphysics is a user-friendly platform for the rigid element analysis and the Multiphysics simulations. It is a very efficient and powerful software which is an FEM based solver. The MEMS comb type accelerometer has been simulated using COMSOL Multiphysics software.

This simulation was intended to find the natural frequency of the proposed mems accelerometer. The simulation model of capacitive comb type accelerometer uses solid mechanics and electromechanics(emi) physics in MEMS module.

The frequency simulation of MEMS accelerometer is shown in figure 5 and is found to be 2.1 kHz.

When an external acceleration of 1g is applied horizontally on x-axis, it causes displacement in the combs. From the simulation results, it is seen that the value of displacement between the combs is 0.05 μm and the value of displacement sensitivity is $5*10^{-9}$m/g as shown in figure 6.

Under steady state an external driving voltage is applied through positive and negative electrodes and simulated to find the Capacitance of MEMS accelerometer. The figure 7 shows the simulation result for the same and was found to be C=0.964 pF

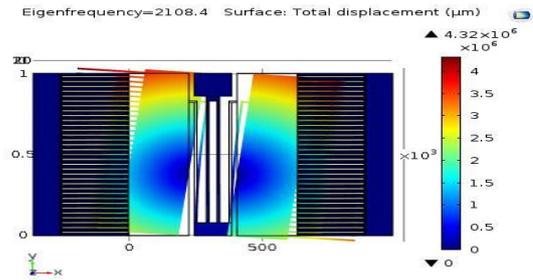

**Fig 5: Frequency simulation of proposed accelerometer.**

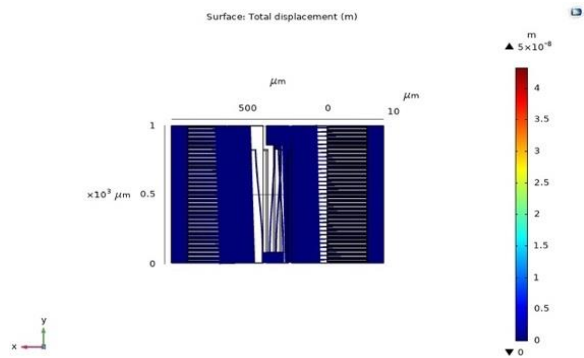

**Fig 6: Displacement sensitivity for the combs is $5*10^{-9}$m/g.**

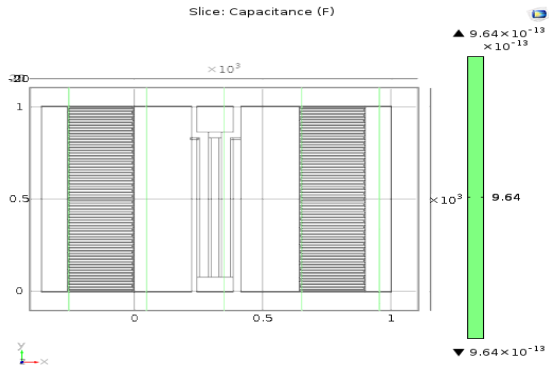

**Fig 7: Capacitance simulation of accelerometer (C=0.964 pF)**

Simulation is done to find the displacement between the sensing fingers for different values of 'g' as shown in the figure 8.

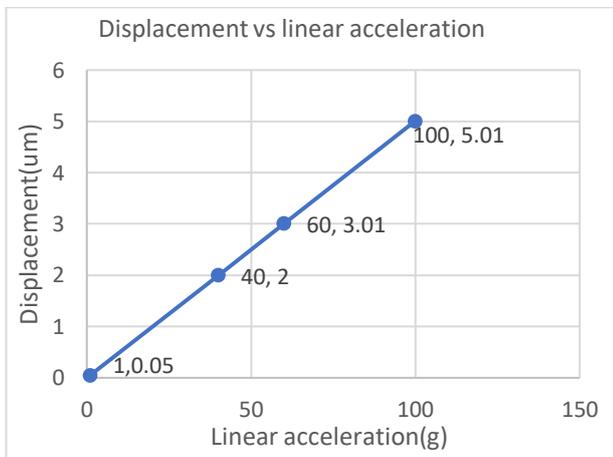

**Fig 8: Variation of displacement vs linear acceleration(g)**

From the above simulation result it is observed that the displacement of proof mass is greater than 5 μm when the acceleration is applied beyond 100g but the minimum distance between the two fingers is 5μm [24]. Hence, the proposed structure cannot be used for acceleration above 100g.

To determine the resonant behaviour of the MEMS accelerometer, AC analysis simulation is performed in MATLAB environment. The figure 9 shows the maximum response and phase angle response as a function of applied external force. It can be observed that the natural frequency of the proposed MEMS accelerometeris 2.6 KHz.

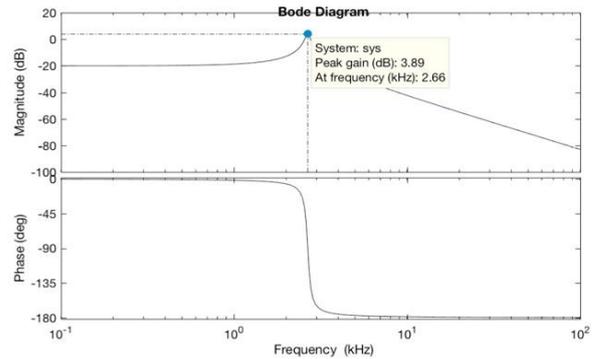

**Fig 9: Frequency and phase response of mems accelerometer**

### VII. OPEN LOOP MODEL OF ACCELEROMETER

In an open-loop system no additional control circuits are required; the electrical output signal is directly interfaced to analogy front end circuit (AFE) for capacitance to voltage conversion and thereafter output voltage is used for measurement analysis [4].

The mathematical equation (8) for open loop accelerometer is represented in a Simulink model as shown in figure 10 [4].

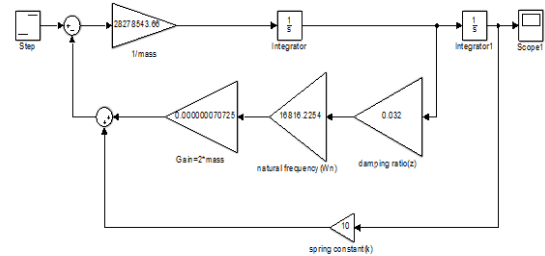

**Fig 10: Simulink model of an open-loop accelerometer**

*A. MatlabSimulation for Open loop Accelerometer:*
The open loop step response is obtained for damping ratio, $\zeta$ =0.03208 (from equation 18) as shown in figure 11a and 11 b.

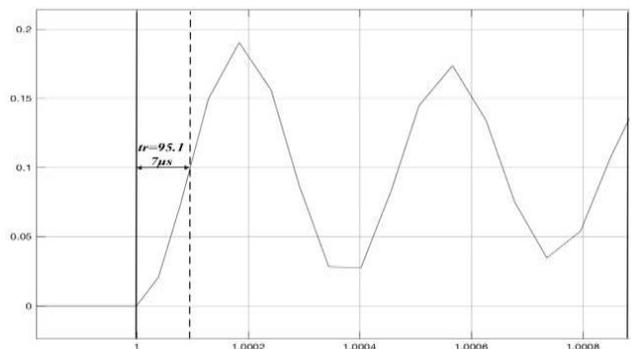

**Fig 11a: Rise time (tr)=95.17 μs when $\zeta$=0.03**

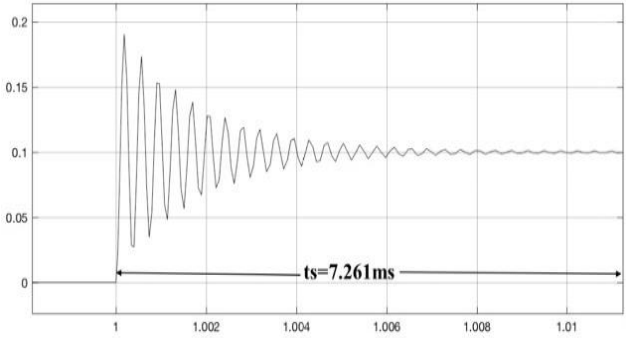

**Fig 11b: Settling time(ts) =7.261 ms when ζ=0.03**

Also, an attempt is made to find the value of ζ for which the system would become stable. From the figures 12 and 13, it is observed that by improving the ζ to 0.5, the system becomes more stable.

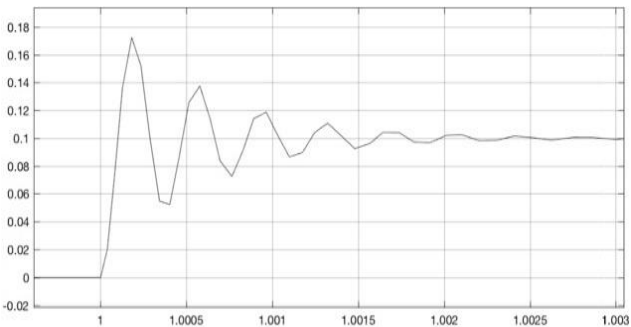

**Fig 12: Step response when ζ=0.1**

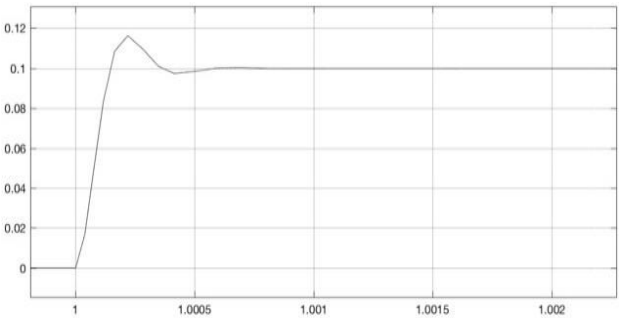

**Fig 13: Step response when ζ=0.5**

### VIII. RESULTS
A single axis MEMS based capacitive accelerometer is designed using COMSOL Multiphysics. Frequency modeling of the proposed accelerometer is done, and the simulations are carried out using COMSOL Multiphysics and MATLAB tools to verify the analytical results obtained. Table III gives the comparison of analytical and simulated results.
The static capacitance and the displacement sensitivity are found and tabulated in the Table III. Displacement analysis and damping ratio analysis are made and hence performance of the device is understood.

**TABLE III: COMPARISON OF ANALYTICAL AND SIMULATED RESULTS**

| Parameters | Analytical values | Simulated values |
|---|---|---|
| Natural frequency ($f_n$) | 2.6 kHz | Comsol: 2.1 kHz Matlab: 2.6 kHz |
| Capacitance($C_0$) | 0.730455 Pf | Comsol: 0.964 Pf |
| Displacement sensitivity ($S_d$) | $3.53625 * 10^{-9}$ m/g | Comsol: $5*10^{-9}$ m/g |
| At ζ =0.03, Rise time($t_r$) and Settling time $t_s$ | $t_r$= 95.36 μs $t_s$= 7.583 ms | Matlab: $t_r$= 95.17 μs $t_s$= 7.261 ms |

It is observed that that the Simulated values are almost nearer to the calculated values as seen from the Table III. However,COMSOL is an FEM based solver whereas MATLAB calculates the behavior of the model as conditions evolve over time or as events occurand each tool may adopt different meshing size and different solving methods which may result in slight difference in the simulated data. Hence the accuracy in the simulation results will vary.

### IX. CONCLUSION
In this work an attempt is made to employ the detailed design,mathematical modelling, sensitivity analysis and simulation of a single axis mems based capacitive accelerometer.
The frequency at which the proposed device resonates is found to be 2.1kHz which is suitable for the applications such as air bag deployment in automobiles.
The displacement analysis presented in this paper concludes that the device should not be operated above 100g as it produces a displacement more than 5μm which in turn may result in short circuit or collision of the comb fingers of the accelerometer.
Analytically, the damping ratio of the proposed accelerometer is found to be 0.03208and concluded this device to be underdamped. It was also observed that by improving the ζ to 0.5, the system becomes more stable. Hence it is decided to extend this work towards improving theζ bychanging the parameters of the proof mass.

### ACKNOWLEDGEMENT
The authors gratefully acknowledge the support extended by Prof Ananth Suresh, Professor, IISc Bangaloreand Dr.Habibuddinshaik, Associate Professor, Physics Department, NMIT, Bangalorein carrying out this work. The authors also thank Mrs.Stuthi, Research Associate,Center for Nano Materials and MEMS,NMIT,Bangalorefor her timely support and the Department of Electrical and Electronics Engineering and authorities of Nitte MeenakshiInstitute ofTechnology,Bangalore for the continuous support and encouragement.
The authors extend their sincere gratitude to Visveswaraya Institute of Technology, Belagavi for the opportunity and support.